\documentclass[9pt,twocolumn,twoside]{opticajnl}
\journal{opticajournal} 

\setboolean{shortarticle}{true}

\usepackage{lineno}

\usepackage{subcaption}
\usepackage{dblfloatfix}
\usepackage{float}
\usepackage{braket}
\usepackage[normalem]{ulem}

\title{A Robust Super-Resolution Classifier by Nonlinear Optics}

\author[1,2]{Ishan Darji}
\author[1,2]{Santosh Kumar}
\author[1,2,*]{Yu-Ping Huang}

\affil[1]{Department of Physics, Stevens Institute of Technology, Hoboken, New Jersey 07030, USA}
\affil[2]{Centre for Quantum Science and Engineering, Stevens Institute of Technology, Hoboken, New Jersey 07030, USA}

\affil[*]{Corresponding Author: yhuang5@stevens.edu}

\begin{abstract}
Spatial-mode projective measurements could achieve super-resolution in remote sensing and imaging, yet their performance is usually sensitive to the parameters of the target scenes. We propose and demonstrate a robust classifier of close-by light sources by using optimized mode projection via nonlinear optics. Contrary to linear-optics based methods using the first few Hermite-Gaussian modes for the projection, here the projection modes are optimally tailored by shaping the pump wave to drive the nonlinear optical process. This minimizes modulation losses and allows high flexibility in designing those modes for robust and efficient measurements. We test this classifier on discriminating one light source and two sources separated well within the Rayleigh limit without prior knowledge of the exact centroid or brightness. Our results show a classification fidelity of over 80\% even when the centroid is misaligned by half the source separation, or when one source is four times stronger than the other.

\end{abstract}

\setboolean{displaycopyright}{false} 

\begin{document}
\maketitle
There have been innumerable attempts to improve the spatial resolution of optical equipment since Lord Rayleigh put forward the empirical diffraction limit \cite{rayleigh_1879,Harris:64,Helstrom:69}. This limit was accepted as the best any optical equipment could do for classical intensity measurements, until the introduction of quantum metrology \cite{haelstorm_thory_1969,haelstorm_res_1973,ram_beyond_2006}, which suggested making use of other properties of light beyond intensity to extract more information.

Tsang et al. \cite{tsang_PRX_2016, tsang_quantum_2019} recently proposed that spatial mode demultiplexing, called ``SPADE'', can beat the diffraction limit. A slew of other methods were proposed and experimentally tested to be more efficient than classical measurements, such as SLIVER \cite{Lu_sliver_2018,zhou_quantum-limited_2019}, SPLICE \cite{tham_beating_2017,Wadood:24}, ROTADE \cite{deAlmeida_misalignmnt_2021}, PSF Shaping \cite{Paur_PSFshaping_2018}, Entangled Partner \cite{sajia_entanglement_2022} in addition to techniques supported by interferometry \cite{Zanforlin_interferometer_2022}, machine learning \cite{pushkina_ML_2021}, statistical post-processing \cite{Bhusal_2022,Tan_QuantSR_2023}, adaptive feedback \cite{tan_adaptive_2022}, and those based on nonlinear optics \cite{Zhang_sr_2020,Zhang:22}. Yet, most of these methods assume ideal laboratory conditions of precisely aligned equipment and sources of equal strength. In real world applications, however, one or both of these conditions would be violated \cite{Schlichtholz:24,Grace:20}. It has also been shown that the quantum advantage disappears if the mode sorter is misaligned from the centroid of the sources \cite{tsang_PRX_2016, deAlmeida_misalignmnt_2021, Sorelli_2021, Grace:20, Grace_2022, tan_adaptive_2022} or if the two sources are of unequal brightness \cite{Linowski_2023,Santamaria:24}.

In this letter, we propose and demonstrate a robust and efficient classifier by extending the nonlinear-optical SPADE (NL-SPADE) to the uses of optimized projection modes \cite{Zhang_sr_2020}. In NL-SPADE, the signals are combined with a pump beam in a nonlinear crystal for sum frequency generation (SFG), difference frequency generation (DFG), or four-wave mixing (FWM)\cite{Boyd_NLOptics}. Using short and tightly focused pump pulses, only signal photons in a single spatial-temporal mode can be converted and subsequently detected \cite{Garikapati_2023}. This implementation of mode projective measurements offers several distinct practical advantages. First, the projection modes are determined by the spatial profile of the pump and the phase matching condition of the nonlinear crystals. Thus there is more freedom to tailor those modes. Second, only the pump profiles are modulated by a spatial light modulator (SLM) before it interacts with the signal. Hence, the incoming signal is left unaltered, avoiding any modulation losses of the original signal. This is important in applications with weak signals, as is usually the case for astronomical observations. Lastly, the nonlinear optics processes can also transduce the signal to another wavelength, such as from near-infrared to visible, where better detectors are available. 

We demonstrate this method experimentally for hypothesis testing of one source versus two sources under the effects of misaligned centroids and unequal brightness separately. Our results find this method to achieve more than 80\% fidelity in classifying the number of sources when they are separated by a fifth of the width of the gaussian point spread function (PSF) of the imaging system, even when the centroid of the sources is misaligned by half of the source separation. For sources of unequal brightness, a fidelity higher than 80\% is obtained again even when one source is four times as bright as the other. These results are recorded with an average number of detected photons as low as 500. The fidelity improves as the photon number increases.

\setcounter{figure}{0}
\begin{figure}[t]
\captionsetup[subfloat]{position=top} 
\centering
\includegraphics[width=0.9\linewidth]{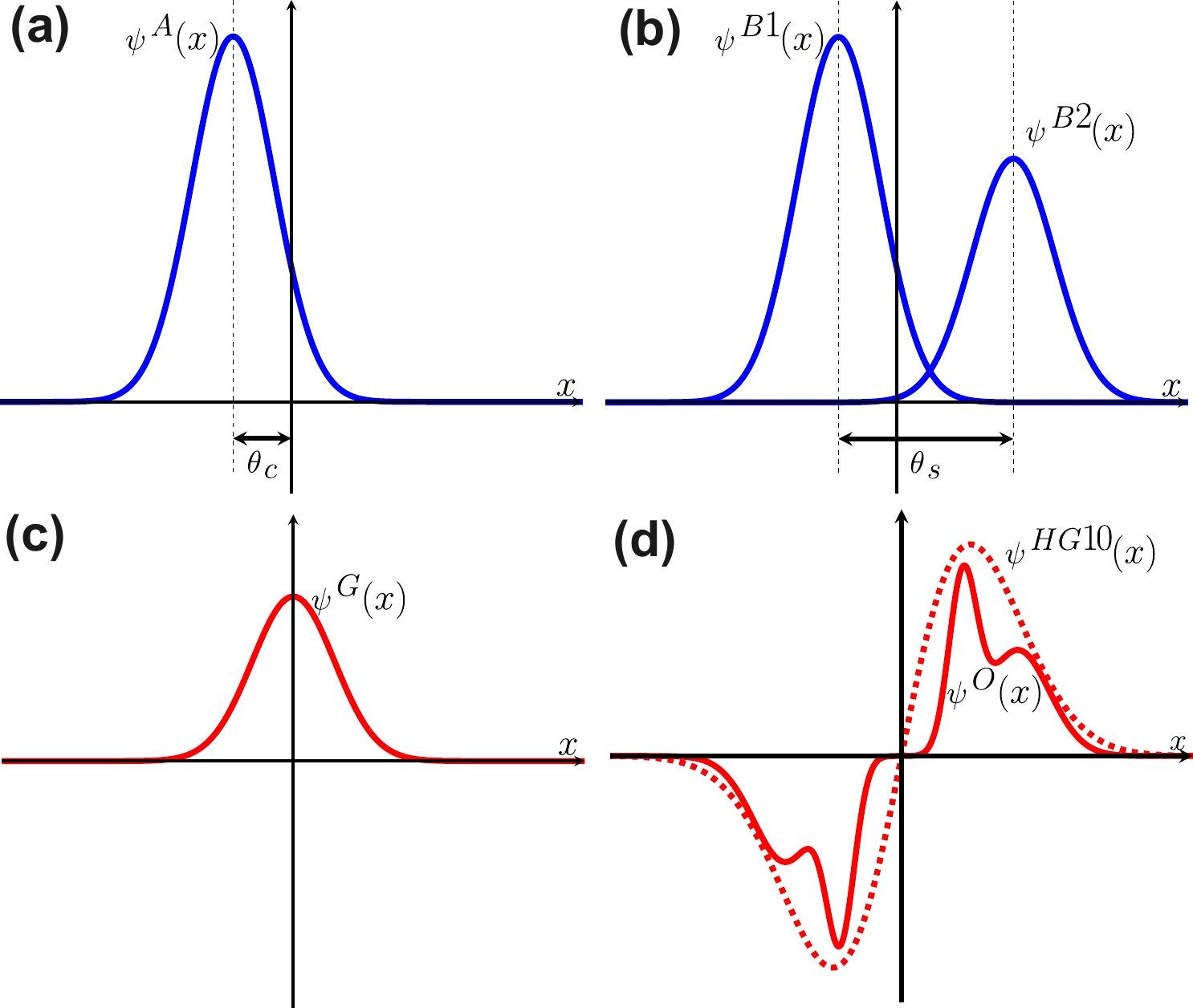}
\caption{One-dimensional representation of spatial modes of: (a) Signal in case A with emission incoming from a single source centered at $\theta_c$ along the x-axis; (b) Signal in case B with emission incoming from two sources separated by $\theta_s$ along the x-axis; (c) Pump in 0$^{th}$ order Gaussian Mode; and (d) Pump in 1$^{st}$ order HG mode (dashed) and in optimized superposition of HG modes (solid).}
\label{fig:wavefunctions}
\end{figure}

We conceptually address the problem of hypothesis testing in the weak signal regime, aiming to distinguish between two scenarios: case A, where the incoming signal is emitted by a single source; and case B, where the incoming signal originates from two sources without spatial coherence. The state of the incoming signals are represented by the following density matrices:
\begin{equation}
\rho_{A}^{} = (1-\eta)\ket{0}\bra{0} + \eta\ket{\psi^{A}}\bra{\psi^{A}}, 
\end{equation}
\begin{equation}
\begin{split}
\rho_{B}^{} &= (1-\eta)\ket{0}\bra{0}  + \eta\epsilon\ket{\psi^{B1}}\bra{\psi^{B1}} \\
&
+ \eta(1-\epsilon)\ket{\psi^{B2}}\bra{\psi^{B2}},\\
\end{split}
\end{equation}
\noindent where $\ket{\psi^{j}}$ is the gaussian PSF, with the superscript $j = A$ denoting the single source for case A, and $j = B1, B2$ denoting the two sources for case B. $\ket{0}\bra{0}$ represents the vacuum state with no incoming photons, $\eta$ is the probability of detecting a photon, and $\epsilon$ is the relative intensity of the sources in case B. 

Figure~\ref{fig:wavefunctions}(a) and (b) visualize the spatial modes of the hypotheses A and B respectively, where the sources are assumed to be separated along the x-axis. The pump modes are centered at the intensity maximum of the signal, whose location can be determined using conventional methods. This is in contrast to other studies \cite{tan_adaptive_2022, Grace:20, Zanforlin_interferometer_2022, Lu_sliver_2018,Wadood:24,Paur_PSFshaping_2018} where the demultiplexing modes are assumed to be aligned with the centroid of the sources, whose location is much harder to be determined in practice. 
In this study, we consider the centroid of this source to be misaligned along the x-axis by $\theta_c$ (see Figure~\ref{fig:wavefunctions}(a)), while $\theta_s$ represents the separation between the two sources in case B (see Figure~\ref{fig:wavefunctions}(b)). 

The incoming signal is combined with the pump inside a nonlinear crystal for sum frequency generation that up-converts the signal photons upon satisfying the phase matching condition with the pump \cite{kumar_upconv_2019, kumar_2021}. The spatial phase profile of the pump is prepared using an SLM, which indirectly tailors the projection modes to achieve the optimal performance. In contrast, for linear optics based SPADE measurements, the projection modes are modulated directly and thus more restrictive, e.g., to certain HG modes. For those using 0th and 1st order HG modes, the resolution advantage quickly disappears when there is centroid misalignment or unequal brightness.

In our nonlinear-optics approach, the first projection mode remains the Gaussian mode (i.e., the 0th order HG mode). However, the second projection mode is a superposition of 30 Hermite-Gaussian (HG) modes \cite{Zhang_sr_2020}. Figure~\ref{fig:wavefunctions}(d) shows an example of the optimized superposition mode as compared with the 1st order HG mode. Let $U_{G}$ and $U_O$ be the operators describing the SFG in the crystal, when pump is in the Gaussian and optimized modes, respectively\cite{Donohue_2015}. At the output of the crystal, the signal becomes $\rho_{j}^{K} = U^{}_{K} \rho^{}_{j} U^{\dag}_{K}$, with $j=A,B$ and $K=G,O$. 

The sum-frequency outputs are then coupled into a single mode fiber with spatial mode $\ket{F}$, before being detected by an avalanche photo diode (APD). The probability of registering an SFG photon is $G_{j} = \bra{F}\rho^{G}_{j}\ket{F}$ for the Gaussian pump mode, and $O_{j} = \bra{F}\rho^{O}_{j}\ket{F}$ for the optimized mode. In experiment, $G_j$ and $O_j$ can be inferred by repeating photon counting measurements. While there are multiple ways of analysis for the hypothesis test, we choose a simple, single-parameter approach by defining an extinction ratio $R_{j} = O_{j}/G_{j}$. By optimizing the pump for the two-source hypothesis, we have $R_{A}<R_{B}$, so that a threshold ratio can be defined as
\begin{equation}
    R_{t} = \frac{1}{2}\left(R_{A} + R_{B} + \Delta R_{A} - \Delta R_{B} \right)
\end{equation}
for the test. Here $\Delta R_{j} = R_{j}\sqrt{\frac{1}{O_{j}}+\frac{1}{G_{j}}}$ accounts for the shot noise uncertainty. If the measured ratio is lower than $R_t$, then the source is classified as case A. Otherwise, it is classified as case B. 

Finally, we define the classification fidelity, as a measure of correct categorization probability,
\begin{equation}
    C = \frac{1}{2} \left[ P(A|A) + P(B|B) \right]
\end{equation}
where $P(j|k)$ ($j,k=A,B$) is the probability of classifying case $k$ to be case $j$. 

\setcounter{figure}{1}
\begin{figure*}[ht]
\centering
\includegraphics[width=0.9\linewidth]{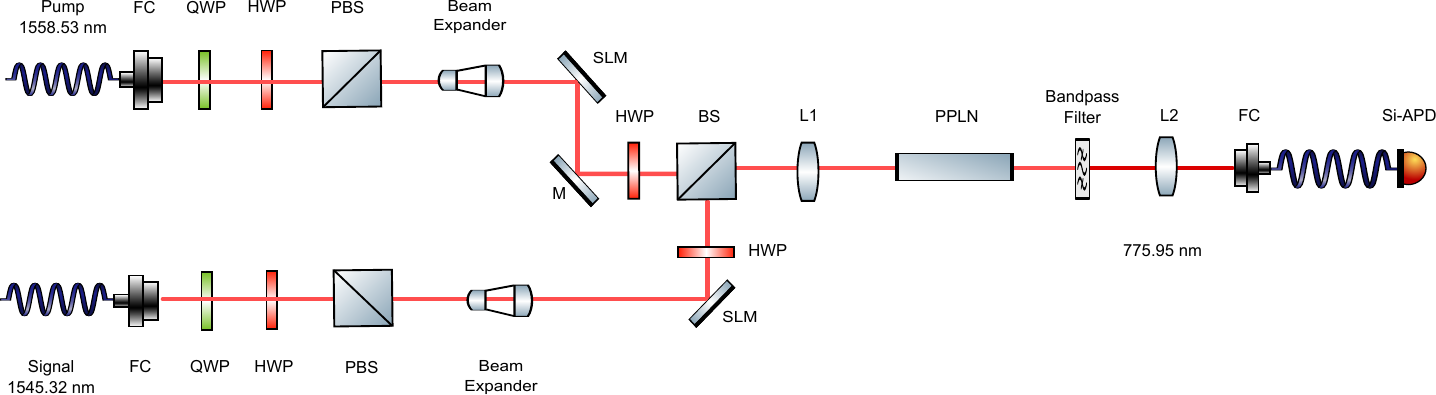}
\caption{Experimental Setup for incoherent source classification beyond the Rayleigh limit using single-pixel mode-selective photon counting technique.  FC - Fiber Coupler, QWP - Quarter Wave Plate, HWP - Half Wave Plate, PBS - Polarizing Beam Splitter, SLM - Spatial Light Modulator, M - Mirror, BS - Beam Splitter, L1/L2 - Lens, PPLN - Periodically Poled Lithium Niabate crystal, Si-APD - Silicone Avalanche Photo-diode.}
\label{fig:exp-setup}
\end{figure*}

The experimental setup is sketched in Figure~\ref{fig:exp-setup}. We use a fiber coupled mode-locked laser (not shown) operating at a 50 MHz repetition rate. Its output is demultiplexed into two beams using two inline, narrow-band wavelength division multiplexers (WDMs) of 0.8 nm linewidth (not shown). The wavelengths of the pump and signal are 1558.53 nm and 1545.32 nm, respectively. They are each coupled into free space, and intensity modulated by sets of quarter-wave plates (QWP), half-wave plates (HWP), and polarizing beamsplitters. Afterwards, their beam sizes are enlarged to about 4 millimeter in full width half maximum (FWHM), before each is incident on an SLM (Santec LCOS SLM-100). The pump is prepared in an optimized superposition of 30 HG modes, while the signal SLM uses blaze grating masks to displace the signal beam on the horizontal axis with micrometer precision\cite{how_to_shape_light}. These beams are then combined at a beam splitter, before being focused using a lens (f=200 mm). At the Fourier plane of this lens, the FWHM beam sizes are approximately 120 µm. A periodically poled lithium niobate (PPLN) crystal (HC Photonics, 50x12.3x1 mm) is placed at the Fourier plane, which facilitates the SFG. The up-converted photons are in the visible spectrum at 775.948 nm wavelength. The bandpass filter eliminates any higher harmonics from the pump and signal, before detection by a Si-APD detector (ID Quantique ID100, dark counts $\sim$2 Hz).

We use two different algorithms in conjunction to optimize the pump mode. This superposition mode consists of thirty $HG_{mn}$ modes, with m = 1,3,5,7,9 along $x$ and n = 0,1,2,3,4,5 along $y$. We start with a simulation of random sampling that prepares a large number of superposition modes whose complex coefficients are distributed randomly. Each of this uniformly distributed set of thirty coefficients is considered as a particle in a higher dimensional hyperspace. We evaluate the photon counts for each of these modes by simulating the SFG process using split-step method\cite{AGRAWAL201327}. This simulation helps us prepare a superposition that can extract the most amount of information for a given scene. Checking this large number of particles directly on the experimental setup is limited by the speed of the SLMs. Hence we run the simulation first to choose 30 superposition pump modes for large $R_B-R_A$ out of 300 randomly prepared modes. Next, those 30 modes are used as the initial guesses for the particle swarm optimization \cite{PSO_2015}, performed in experiment. We run twenty iterations of the algorithm for each optimization run. The convergence criteria is defined to maximize the difference of the ratios, $R_B-R_A$. Additionally, we reduce the weights of the particle velocities by one order of magnitude every third iteration, to improve the rate of convergence. In this optimization, there is no misalignment or unequal brightness assumed. Yet, as shown below, high performance is achieved for a large range of misalignment and brightness inequality, which speaks to the robustness and applicability of our method. 

\begin{figure}[ht]
\captionsetup[subfloat]{position=top}  
\centering
\includegraphics[width=0.9\linewidth]{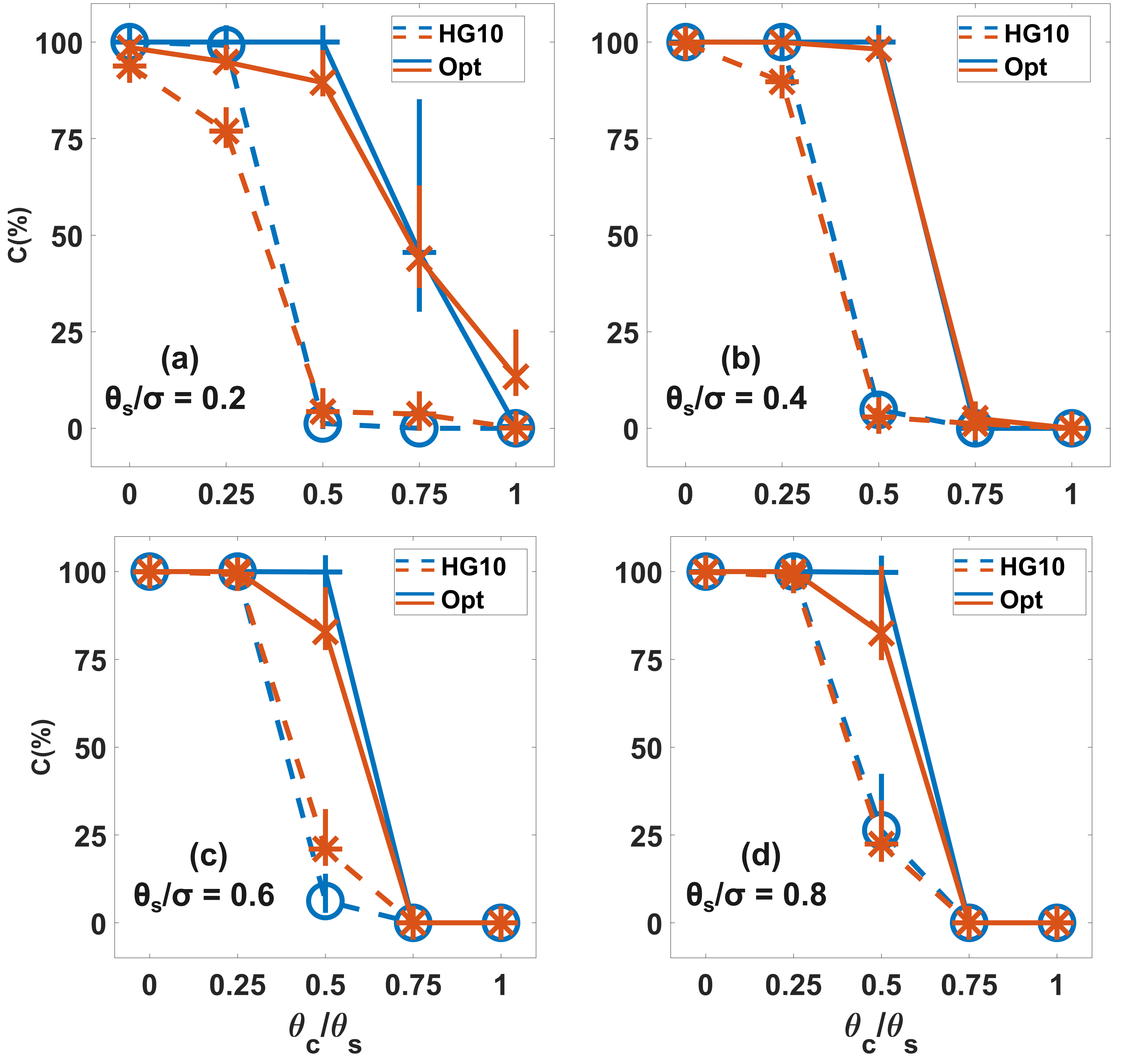}
\caption{Classification fidelity (C) as a function of misalignment ($\theta_c$) for four different source separations, with $\theta_s/\sigma$ = (a) 0.2; (b) 0.4; (c) 0.6; and (d) 0.8. For each case the sources are of equal brightness ($\epsilon = 0.5$). The signal intensity is modulated while the pump is unchanged such that average numbers of detected photons are about 500 (in orange) and 5000 (in blue). The error bars represent 95\% confidence intervals from 10 different runs of the experiment, with 1000 data points in each run.}
\label{fig:misalignment_fidelity}
\end{figure}

We test four different source separations, each with misalignment varied up to the source separation (i.e., $\theta_c\in [0,\theta_s]$). Figure~\ref{fig:misalignment_fidelity} shows the impact of misalignment on the system’s classification fidelity. Blue and orange curves represent the results when the average detected photons were about 5000 and 500, respectively. Note that the two sources of case B have equal strength ($\epsilon = 0.5$) for all results in Figure~\ref{fig:misalignment_fidelity}. The optimized modes (solid curves) are noticeably more efficient at discriminating the sources as compared to only the first order HG mode (dashed curves).

For higher average photon counts (see the blue curves in Figure~\ref{fig:misalignment_fidelity}), the classification fidelity of the optimized modes is maintained at 100\% for misalignment up to 0.5$\theta_s$, whereas the fidelity with the first order HG mode drops to nearly 0\% at this misalignment. For larger misalignment,  the fidelity drops quickly. For lower average photon counts (see the orange curves in Figure~\ref{fig:misalignment_fidelity}), the advantage of using the optimized mode is maintained. With the HG10 mode, the fidelity drops to 0\% again as the misalignment increases up to 0.5$\theta_s$, while the optimized modes maintain their fidelity above 80\%. Our results show that with an optimized pump, the classification fidelity does not degrade even with 500 photon counts. As we further decrease the photon counts, the classification fidelity drops as the shot noise dominates \cite{Zhang_sr_2020}.

\begin{figure}[ht]
\captionsetup[subfloat]{position=top}  
\centering
\includegraphics[width=0.99\linewidth]{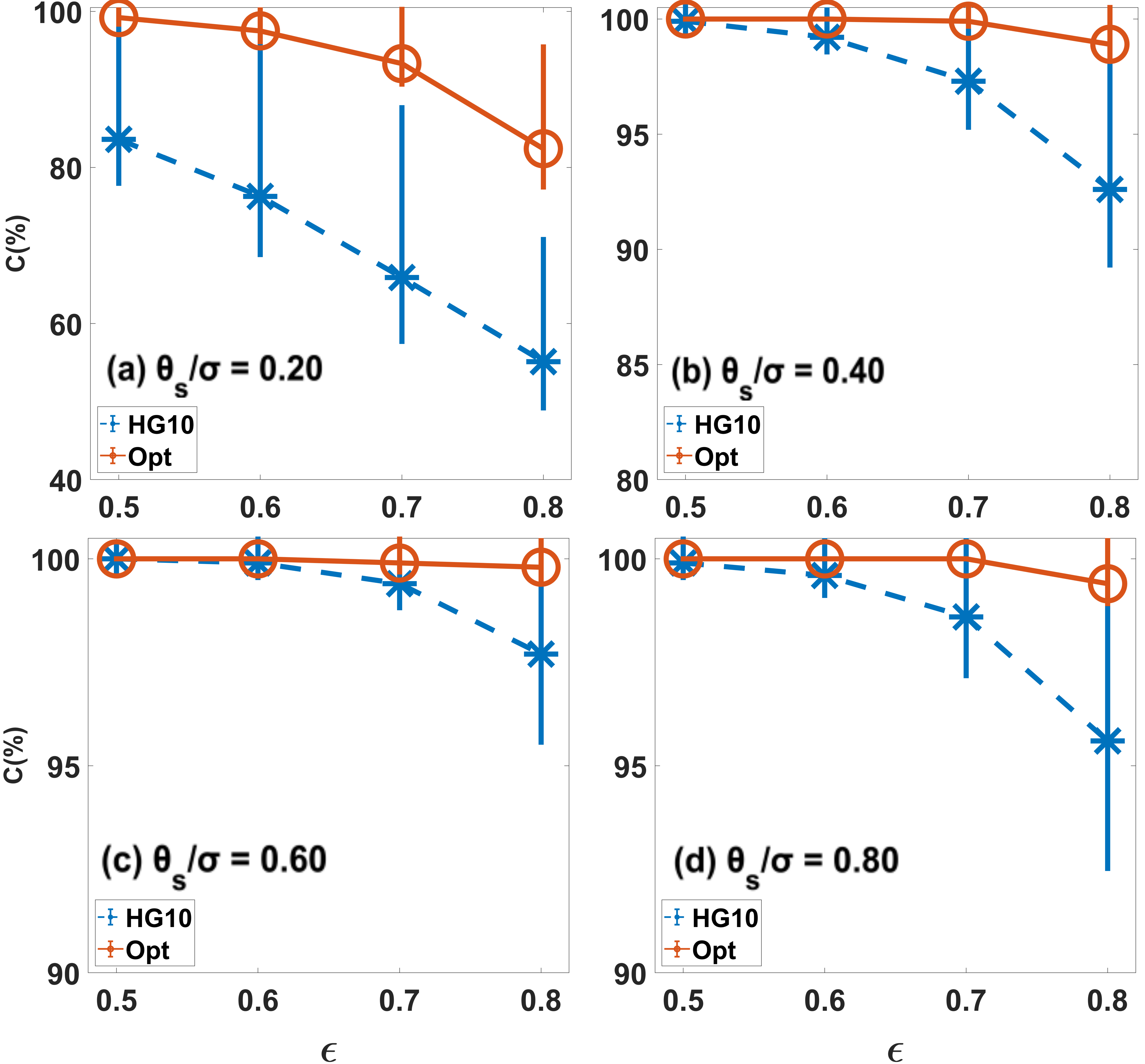}
\caption{Classification fidelity (C) as a function of relative brightness ($\epsilon$) for four different source separations, with $\theta_s/\sigma$ = (a) 0.2; (b) 0.4; (c) 0.6 and (d) 0.8. The sources are not misaligned ($\theta_c = 0$). The average number of detected photons is about 750. The error bars represent 95\% confidence intervals from 10 different runs of the experiment, with 100 data points in each run.}
\label{fig:brightness_fidelity}
\end{figure}

It was shown that the error probability (which is the complement of the fidelity defined here) increases as the source separation decreases \cite{Wadood:24} for linear SPADE. This is reflected by the dashed curves at 0 misalignment in Figure \ref{fig:misalignment_fidelity}. However, by using optimized mode, these error probabilities are reduced and even maintained under misalignment.

We also test for four different relative brightness ($\epsilon \in [0.5, 0.8]$) for similar source separations. In Figure~\ref{fig:brightness_fidelity}, we see the impact of unequal brightness on classification fidelity. For each case, 750 photons are detected on average. The optimized modes provide a fidelity that is above 80\% for separation as small as 0.2$\sigma$ (see the orange curve in Figure~\ref{fig:brightness_fidelity}(a)) when one source was four times stronger than the other in case B. On the other hand, the fidelity of the HG10 mode is about 50\% (blue curve in Figure~\ref{fig:brightness_fidelity}(a)) for this scene. At larger separations (Figure~\ref{fig:brightness_fidelity}(c) and (d)), the fidelity starts to deteriorate when using only HG10 mode (in blue) as the brightness mismatch increases, but it is unhindered close to 100\% when using the optimized modes (see the orange curves).

We also note that for a given pump wavelength, the current experimental setup is efficient for a narrow band of signal spectrum. Broadband sum-frequency generation methods can be implemented to improve the efficiency of this technique over a wider signal spectrum.

In summary, we explored the nonlinear-optical spatial-mode demultiplexing with optimized mode projection for practical applications in presence of centroid misalignment and unequal brightness. Our results find superior performance to linear-optical SPADE, in terms of robustness, flexibility, detection efficiency and noise. Such advantages are established with only simple and quick optimization. Even higher performance is expected with more exhaustive optimization. Because of the flexibility provided by the nonlinear-optical implementation, this method could be extended to complex scenes with many sources. It holds promise for applications in astronomy, remote sensing, and biomedical imaging. 

\section*{Funding} 
This research was supported in part by NASA under grant number 80NSSC22K1918.
\section*{Disclosures} The authors declare no conflicts of interest.
\section*{Acknowledgments} 
We acknowledge Brian Baker from Ball Aerospace \& Technologies for some inspirational discussions. 






\section*{Data availability} The data, and result analysis presented in this Letter are publicly available in the GitHub repository \url{ https://github.com/idarji-stevens-edu/NL-Spade.git}.


\bibliography{sample}

\bibliographyfullrefs{sample}



\end{document}